\documentstyle[preprint,prb,aps]{revtex}
\begin{document}
\draft
\title{Driven Diffusion in the Two-Dimensional Lattice Coulomb Gas;
A Model for Flux Flow in Superconducting Networks}
\author{Jong-Rim Lee}
\address{Center for Theoretical Physics, Seoul National University,
Seoul 151-742, Republic of Korea}
\author{S. Teitel}
\address{Department of Physics and Astronomy, University of Rochester,
Rochester, NY 14627}
\date{\today}
\maketitle
\begin{abstract}
We carry out driven diffusion Monte Carlo simulations of the
two dimensional classical lattice Coulomb gas in an applied
uniform electric field, as a model for
vortex motion due to an applied d.c. current in a periodic
superconducting network.  A finite-size version of dynamic
scaling is used to extract the dynamic critical exponent $z$,
and infer the non-linear response at the transition temperature.
We consider the Coloumb gases $f=0$, and $f=1/2$, corresponding to a
superconducting network with an applied transverse magnetic field
of  zero, and one half flux quantum per unit cell, respectively.
\end{abstract}

\narrowtext
\section{Introduction}

Phase transitions in two dimensional ($2d$)
superconducting networks, such as periodic Josephson
junction arrays and superconducting wire nets, has been a topic of much
investigation.\cite{topic}
Theoretically, the phase transitions in such systems have
been most extensively studied by $equilibrium$ calculations and
simulations.\cite{theory,f12,Grest,LeeCG,Leef12,Jose}
Experimentally however, it has been most common to measure steady-state
current-voltage
($I-V$) characteristics, and look for a cross-over from linear to non-linear
resistivity as a signal of the superconducting
transition.\cite{expt,expt2}  In this
regard, the Kosterlitz-Thouless (KT) model\cite{kt} of a vortex pair
unbinding transition
makes a clear prediction:\cite{iv,iv2} as one heats up through the transition
temperature $T_{KT}$, the $I-V$ characteristics should
make a discontinuous
change from $V\sim I^3$ exactly at $T_{KT}$ to $V\sim I$ above $T_{KT}$.
In experimental studies of superconducting $2d$ films\cite{Min} and
networks\cite{expt2}
however, as well as in numerical simulations,\cite{Mon}
agreement with this prediction has been claimed in some cases,
not found in others, and is ambiguous in yet others.  In particular,
it is not clear how
this prediction may become modified when a transverse magnetic field is
applied to the sample.  In this case, the melting of the ground state
vortex lattice induced by the magnetic field, may change the universality
class of the superconducting transition, and lead to different steady
state behavior.
In view of these questions, it remains of interest to establish what
steady state $I-V$ behavior may be expected at criticality, for specific
simple cases.

Recently, a new dynamic scaling conjecture was proposed
by Fisher {\it et al.}\cite{Fis}
to describe $I-V$ characteristics in superconducting
systems.  Although this approach was developed for application to the
``vortex-glass" transition in high temperature superconductors, it should
apply equally well to any superconducting transition which is believed to
be second order. In this work we carry out steady-state ``driven diffusion"
Monte Carlo simulations\cite{driven}
of the $2d$ lattice Coulomb gas\cite{Grest,LeeCG} in a uniform applied
electric field, as a model for vortex motion due to a uniform applied
d.c. current, in a periodic superconducting network.  We consider the
special cases of $f=0$, corresponding to no transverse magnetic field,
and $f=1/2$, corresponding to a transverse magnetic field of one half
flux quantum per unit cell of the network.  We apply a finite-size version
of the new dynamic scaling conjecture to analyze our data, and extract the
dynamic critical exponent $z$, and the power law of the $I-V$ characteristic
at the superconducting transition.  For $f=0$, we find $z=2$, consistent
with the Kosterlitz-Thouless prediction.  For $f=1/2$, we again find
$z=2$, consistent with the KT model, but in disagreement with expectations
from recent equilibrium simulations of this model.

The remainder of this paper is organized as follows.  In Section II
we give the theoretical framework for our simulations.  We present our
Coulomb gas model and the driven diffusion Monte Carlo algorithm.
We review the KT vortex pair unbinding prediction,
and we describe the finite-size version of dynamic scaling.
In Section III we present our numerical results for $f=0$ and
for $f=1/2$.  In Section IV we give our conclusions.

\section{Theoretical Framework}

\subsection{The Driven Diffusive Lattice Coulomb Gas}

The standard model\cite{theory,f12}
to describe behavior in a $2d$ superconducting
network, is the uniformly frustrated $2d$ XY model, given by the
Hamiltonian,
\begin{equation}
{\cal H}_{XY} = \sum_{\langle ij\rangle}U(\theta_i-\theta_j-A_{ij}).
\label{eq:hxy}
\end{equation}
Here $\theta_i$ is the phase of the superconducting wavefunction at
node $i$ of the periodic network, the sum is over all nearest neighbor
bonds $\langle ij\rangle$ of the network, and
\begin{equation}
A_{ij}={2e\over \hbar c}\int_i^j\vec A\cdot d\vec l
\label{eq:aij}
\end{equation}
is the line integral of the magnetic vector potential $\vec A$
across the bond from $i$ to $j$.  For a uniform magnetic field
applied transverse to the plane of the network, the sum of the
$A_{ij}$ around (going counter-clockwise) any unit cell is,
\begin{equation}
\sum_{cell} A_{ij} = 2\pi f,\qquad\qquad f=B{\cal A}/\Phi_0,
\label{eq:f}
\end{equation}
where ${\cal A}$ is the area of a unit cell, and the constant
$f$ is the density of magnetic flux quanta ($\Phi_0 = 2e/hc$) per unit cell.
$f$ is referred to as the ``uniform frustration".
$U(\phi)$ is the interaction potential between the nodes of the network,
which is periodic in $\phi$ with period $2\pi$, and has a minimum at
$\phi=0$.
For a Josephson junction array, one takes\cite{Tink}
$U(\phi)=-J_0\cos (\phi)$.
For a wire net, the Villain,\cite{Vil} or ``periodic Gaussian"
interaction may be more appropriate.\cite{TJwire}

It is generally believed that it is the excitation of vortices
in the superconducting phase $\theta_i$, that is responsible for
the superconducting transition in such networks.\cite{theory}
Since $2d$ vortices interact with a logarithmic potential,\cite{kt}
the Hamiltonain (\ref{eq:hxy}) is assumed to be in the same
universality class (for the Villain interaction,\cite{Vil} the mapping
via duality, is exact\cite{dual})
as the following Hamiltonian for Coulomb interacting point charges,
\begin{equation}
{\cal H}_{CG}={1\over 2}\sum_{ij}(n_i-f)V(\vec r_i-\vec r_j)(n_i-f).
\label{eq:hcg}
\end{equation}
Here, $i$ and $j$ label the $dual$ sites of the original superconducting
network, $n_i=0,\pm 1,\pm 2,...$ is the integer vorticity of the
superconducting phase $\theta $ at site $i$,
and $V(\vec r)$ is the $2d$ lattice Green's function, which satisfies,
\begin{equation}
\Delta^2 V(\vec r_i-\vec r_j) = -2\pi\delta_{ij}
\label{eq:laplace}
\end{equation}
where $\Delta^2$ is the discrete Laplacian.
In this work we restrict our interest to a square network,
with periodic boundary conditions.  In this case, $V(\vec r)$
is explicitly given by
\begin{equation}
V(\vec r)={1\over N}\sum_k e^{i\vec k\cdot\vec r}
{\pi\over 2-\cos k_x-\cos k_y}
\label{eq:vk}
\end{equation}
where $\vec k$ are the allowed wave vectors with $k_\mu=(2\pi n_\mu/L)$,
with $n_\mu =0,1,...,L-1$.  $L$ is the length of the network, and $N=L^2$.
For large $\vec r$,
\begin{equation}
V(\vec r)\sim \ln r.
\label{eq:log}
\end{equation}
Since $V(\vec r=0)$ is divergent, the partition sum over
$\{ n_i\}$ is restricted to neutral configurations where
\begin{equation}
{1\over N}\sum_i n_i = f.
\label{eq:neutrality}
\end{equation}
See Ref.\onlinecite{LeeCG} for further details.  Thus the average
density of vortices is equal to the density of flux quantum of the
applied magnetic field.

The Hamiltonian (\ref{eq:hcg}) therefore represents a density $f$ of integer
point charges, on a uniform compensating background charge $-f$,
interacting with the $2d$ Coulomb potential.  For $f=0$, the ground state
is the vacuum, and low lying excitations are bound neutral pairs of
$n_i=\pm 1$.  For $f=1/2$, the ground state is an ordered checkerboard
lattice of charges\cite{f12} with $n_i=+1$ on every other site, as shown
in Fig.\ 1.  Low lying excitations
may be viewed either as a displacement of one of the charges in the ground
state lattice, or equivalently the creation of a neutral $\Delta
n_i=\pm 1$ pair
of charges superimposed on the ground state lattice configuration.  It is
this Coulomb gas analog of the superconducting network which we will use to
carry out our simulations.

To model flux flow resistance in the superconducting network,
we apply a uniform electric field to the Coulomb gas charges $n_i$; this
models the uniform Magnus force\cite{Tink}
that an applied d.c. current exerts
on vortices in the superconductor.  The net charge current density in the
Coulomb gas then corresponds to the net vortex current density in the
superconductor, which is proportional to the net flux flow electric
field (or voltage drop per unit length) in the superconductor.
If we denote as $\vec E$ the applied electric field,
and $\vec J$ the resulting charge current density, in
the Coulomb gas analog, and ${\cal \vec J}$ the applied d.c. current density,
and ${\cal \vec E}$ the resulting flux flow electric field,
in the superconducting network, then the correspondence between the models
is given by
\begin{equation}
\vec E\leftrightarrow {\cal \vec J},\qquad\qquad \vec
J\leftrightarrow {\cal \vec E}.
\label{eq:corresp}
\end{equation}

The simulation of the Coulomb gas in the presence of the uniform $\vec E$ is
carried out by use of the ``driven diffusion" technique.\cite{driven}
We imagine adding to the Hamiltonian (\ref{eq:hcg}) the term
\begin{equation}
\delta {\cal H}=-\sum_in_i\vec r_i\cdot\vec E
\end{equation}
which represents a dipole interaction between the charges $n_i$ and
the electric field $\vec E$.
Although ${\cal H}_{CG}+\delta {\cal H}$ is unbounded,
and therefore not a valid
Hamiltonian in a global sense, when it is used
as a $local$ energy function for computing energy differences,
in connection with the standard Metropolis
Monte Carlo algorithm for accepting or rejecting proposed excitations,
it yields an enhanced probability (consistent with detailed balance)
for accepting excitations with a net
movement of charge in the direction of $\vec E$,
thus setting up a non-equilibrium steady-state with a finite charge current
density $\vec J$ flowing parallel to $\vec E$.

In our simulations, we have chosen $\vec E=E\hat y$, along one of
the axes of the periodic lattice of sites.  At each step of the simulation
we pick at random a pair of nearest neighbor sites $(i_0,i_1)$.
For the $f=0$ case (where the ground state is $n_i=0$),
we add a positive unit charge to site $i_0$, ie.
$\Delta n_{i_0}=+1$, and a negative unit charge to site $i_1$, ie.
$\Delta n_{i_1}=-1$.  For the $f=1/2$ case (where the ground state
is as in Fig.\ 1),
we simply interchange the
charges, $n_{i_0}$ and $n_{i_1}$, at the two sites. For $f=1/2$, this
restricts configurations to those where half of the $n_i$ are $+1$ and
the other half are $0$; charges
$n_i=-1$ or $+2$ are not allowed.  Tests showed that these other values
of $n_i$ correspond to higher energy excitations, which are negligible
at the temperatures of interest, ie. the melting temperature of the
ground state charge lattice.
In both the $f=0$ and the $f=1/2$ cases, the change in energy
due to the addition of the excitation is computed using
${\cal H}_{CG}+\delta {\cal H}$, and the excitation is accepted
or rejected using the Metropolis algorithm. In both cases, acceptance
of the excitation gives a contribution to the average current density,
\begin{equation}
\Delta \vec J = \Delta n_{i_0}{\vec r_{i_0}-\vec r_{i_1}\over 2}
+\Delta n_{i_1}{\vec r_{i_1}-\vec r_{i_0}\over 2}=\Delta n_{i_0}(
\vec r_{i_0}-\vec r_{i_1})
\label{eq:delj}
\end{equation}
where $\Delta n_i$ is the change in $n_i$ at site $i$ created by adding
the excitation, and our algorithm always satisfies $\Delta n_{i_0}=-
\Delta n_{i_1}$.

While the above driven diffusion Monte Carlo algorithm encodes a specific
dynamics, which in detail may well be different from the true
microscopic dynamics of vortices in superconductors, our hope is that
qualitative behaviors which are largely determined by energetics,
particularly the non-linear form at criticality, will be preserved.
We have chosen\cite{WW} to simulate the driven diffusion Coulomb gas, rather
than the more realistic RSJ dynamics for an XY model of a Josephson
array,\cite{Mon,jjdy} because in the Coulomb gas algorithm one directly
moves the important degrees of freedom, the positions of the vortices.
This results in a computationally faster algorithm for several reasons:
($i$) spin wave degrees of freedom are eliminated; ($ii$) the effective
energy barrier for an isolated vortex to hop to a neighboring cell in
the XY model\cite{iv2} is removed, since in the Coulomb gas this
hop takes place in one discrete step; ($iii$) the RSJ dynamics requires
a computation of order $L^2$
(or $L\ln L$ for improved algorithms\cite{jjdy2})
at each step of simulation, where the Coulomb gas requires a computation
of order $L^2$ only when the trial excitation is accepted;\cite{LeeCG}
for the low acceptance ratios we find, this effect is significant.

$N=L^2$ steps of the above updating process will be referred
to a one MC pass.  In our runs, presented in Section III, an initial
10,000 passes were typically discarded to equilibrate the system.
Following this equilibration, five independent runs (using different
random number sequences) of 200,000 passes each,
were used to compute averages.  Our error bars are estimated from the
standard deviation of the averages from these five runs.

\subsection{Kosterlitz-Thouless Pair Unbinding Model}
\label{ktsec}

We now review the Kosterlitz-Thouless model of pair unbinding,\cite{kt,Min}
as applied
to the determination of non-linear steady-state behavior\cite{iv} below the
transition temperature $T_{KT}$.  If we consider the addition of a neutral
pair of charges $\Delta n_{i}=\pm 1$ separated by a distance $\vec r$,
we may estimate the free energy of this pair in the presence of
all other charges as,
\begin{equation}
F_{pair}(\vec r) ={1\over\epsilon}(\ln |\vec r|-\vec E\cdot\vec r).
\label{eq:Fpair}
\end{equation}
Here $\epsilon$ is the effective long wavelength dielectric function
of the Coulomb gas, which serves to screen the logarithmic interaction
between the members of the pair, as well as to screen the interaction of
the pair with the external field $\vec E$.  A pair oriented along the
direction of $\vec E$ therefore sees a potential maximum at $r_c=1/E$.
If the pair is able to overcome this potential barrier through thermal
fluctuations, the pair can then lower its free energy by
increasing the separation $r$
without bound.  The pair will thus unbind, and give
a contribution to a net current of charges flowing along the direction
of $\vec E$.  The rate for such critical pair unbindings to occur is
given by the Boltzmann factor,
\begin{equation}
W_{pair}\sim e^{-F_{pair}(r_c)/T}\sim E^{1/\epsilon T}.
\label{eq:W}
\end{equation}
Such critical pairs will expand until they recombine with other such
free charges into new bound pairs.  This unbinding and recombination
of pairs leads\cite{iv} to an effective density of free charges $n_f$,
\begin{equation}
n_f\sim (W_{pair})^{1/2}.
\label{eq:nfree}
\end{equation}
Using Eqs.(\ref{eq:W},\ref{eq:nfree}), the net current that flows due to
pair unbinding is then,
\begin{equation}
J\sim n_f E\sim E^{1+1/2\epsilon T}.
\label{eq:je}
\end{equation}
The insulator to metal transition in the Coulomb gas, where $\epsilon$
diverges, marks the cross over from non-linear to linear
$J-E$ characteristics.  Using the correspondence of Eq.(\ref{eq:corresp}),
together with $I=L{\cal J}$ and $V=L{\cal E}$ for the total current
and total voltage drop in a superconducting network, we see that
this Coulomb gas insulator to metal transition corresponds to the
superconducting to normal transition in the superconducting network.

For $f=0$, where the ground state is the vacuum, pair unbinding
excitations such as above, are believed to be the only source of
net charge current.  The Kosterlitz-Thouless model is
expected to describe the insulator to metal transition in this system,
and gives the prediction\cite{kt,Min} that $\epsilon^{-1}(T)$ has a universal
discontinuous jump to zero
exactly at the transition
temperature $T_{KT}$, with,
\begin{equation}
1/\epsilon(T_{KT})T_{KT}=4.
\label{eq:jump}
\end{equation}
Thus, exactly at $T_{KT}$, Eq.(\ref{eq:je}) gives the non-linear
behavior, $J\sim E^3$.  The corresponding result in the superconducting
network is $V\sim I^3$.

For $f=1/2$, where the ground state is the doubly degenerate
ordered charge lattice
shown in Fig.\ 1, the above pair unbinding continues to provide
a mechanism for non-linear response below the insulator to metal
transition temperature, which we will continue to refer to as $T_{KT}$.
However there are now also other possible excitations, involving the
formation of domains of oppositely oriented ground state, which might
possibly contribute\cite{Mon} to a non-linear response in $J$.  Thus no clear
prediction exists for the form of the non-linear response at the
transition.

Similarly, the nature of the $equilibrium$ transition in the $f=1/2$
model remains controversial.\cite{f12,Grest,Leef12,Jose}
If $T_{KT}$ is the insulator to
metal transition, the Kosterlitz-Thouless arguments continue to provide
a lower bound on a discontinuous jump in $\epsilon^{-1}$, ie.
$1/\epsilon(T_{KT})T_{KT}\ge 4$.  It is unclear however, whether this
is satisfied only as an inequality, or whether the universal
KT behavior as in Eq.(\ref{eq:jump}) continues to hold.  Additionally,
there is a well defined temperature $T_M$ in the model, corresponding
to the melting of the ordered ground state charge lattice.  General
arguments\cite{Nien} suggest the bound $T_M\ge T_{KT}$,
however it remains unclear
whether or not these two temperatures are in fact equal.
It is further unclear whether the melting
transition at $T_M$ is Ising-like (as the double discrete symmetry
of the ground state suggests), or whether the long range
interactions between the charges cause the melting to fall in a new
universality class.  Our present work was in part motivated by the
hope that dynamic calculations might shed some light on
these remaining equilibrium questions.

\subsection{Finite-Size Dynamic Scaling}

Recently, Fisher {\it et al.}\cite{Fis}
have proposed the following dynamic scaling
relation, for an infinite superconducting
system with a second order phase transition
at $T_c$.  The relation between the dissipative electric field
${\cal E}$, and the applied d.c. current density ${\cal J}$, is given by,
\begin{equation}
{\cal E}={\cal J}\xi^{d-2-z}\Phi_\pm({\cal J}\xi^{d-1}/T)
\label{eq:dscal}
\end{equation}
where $\xi$ is the spatial correlation length, $d$ the dimensionality
of the system, $z$ the dynamic scaling exponent, and $\Phi_\pm$ the
scaling function above and below the transition.  The most natural
generalization of this form, to a system of finite length $L$, is,
\begin{equation}
{\cal E}={\cal J}b^{d-2-z}\Phi({\cal J}b^{d-1}/T,tb^{1/\nu},b/L)
\label{eq:fdscal}
\end{equation}
where $b$ is an arbitrary length rescaling factor, $t=(T-T_c)/T_c$,
and $\nu$ is the correlation length scaling exponent, $\xi\sim t^{-\nu}$.
The form Eq.(\ref{eq:dscal}) can be obtained from
Eq.(\ref{eq:fdscal}) by choosing $b=\xi$, and taking $L\to\infty$.
For finite $L$, $\Phi$ is a continuous function as $t$ passes through
zero.
Only in the $L\to\infty$ limit does $\Phi({\cal J},t,0)$ become
discontinuous as $t$ passes through zero; this determines the
different scaling functions $\Phi_+$ and $\Phi_-$ of Eq.(\ref{eq:dscal}).

{}From Eq.(\ref{eq:fdscal}) one can now determine the scaling behavior
at criticality, $t=0$.  Choosing $b={\cal J}^{-1/(d-1)}$, $L\to\infty$,
and setting $t=0$, one finds,
\begin{equation}
{\cal E}={\cal J}^{1-(d-2-z)/(d-1)}\Phi(1/T_c,0,0)\sim {\cal J}^{(1+z)/(d-1)},
\label{eq:ivtc}
\end{equation}
as found by Fisher {\it et al}.\cite{Fis}
Thus at $T_c$, one always expects a
non-linear power law response.  For $d=2$, a dynamic exponent of $z=2$
recovers the ${\cal E}\sim {\cal J}^3$ result of the KT pair unbinding
picture.

Using the correspondences of Eq.(\ref{eq:corresp}), we now recast the
scaling equation (\ref{eq:fdscal}) into a form for use with our Coulomb gas
model.  Choosing the rescaling length $b=L$, we get,\cite{scaling}
\begin{equation}
J=EL^{d-2-z}\Phi(EL^{d-1}/T,tL^{1/\nu},1).
\label{eq:bel}
\end{equation}
Finally, for $d=2$, exactly at criticality, $t=0$, we have the scaling,
\begin{equation}
J=EL^{-z}\Phi (EL/T_c,0,1).
\label{eq:tcscal}
\end{equation}
To fit this scaling equation to our Monte Carlo data, and extract the
dynamic exponent $z$, we use the method used by
Nightingale and Bl\"ote\cite{night}
for similar equilibrium problems.  We consider behavior exactly at $T_c$,
as a function of varying $E$, in the limit of large $L$
but small $EL$.  Expanding the scaling function $\Phi$ gives,
\begin{equation}
J(E,L)=EL^{-z}[\Phi_0+\Phi_1EL+\Phi_2(EL)^2+O(EL)^3].
\label{eq:expscal}
\end{equation}
Truncating this expansion at any finite order, we perform a
least $\chi^2$ non-linear fit\cite{numr} of our Monte Carlo
data to Eq.(\ref{eq:expscal}) to determine the unknown parameters
$z$, and the $\Phi_i$.

We check for stability of our fit by increasing the
order of the expansion, and by dropping data from successively smaller
values of $L$, and checking if the fitted parameters change within
our estimated statistical error.  Statistical errors in the fitted
parameters are estimated
by generating many synthetic data sets, by adding random noise to each
of the original MC data point.  The noise for each data point
is taken from a Gaussian distribution whose width is set equal to the
estimated statistical error of the original MC data point.  Using these
fictitious data sets, we repeat the fitting procedure to obtain new
fitted parameters.  The estimated error of a parameter is then taken as
the standard deviation of the results from all the fictitious data sets.

\section{Numerical Results}

\subsection{$f=0$}

For our simulations of the $f=0$ Coulomb gas, corresponding to the
ordinary XY model, we use as the equilibrium KT transition temperature
$T_{KT}=0.218$, as determined by one of us\cite{Leeth} from
a finite size scaling analysis applied
to equilibrium simulations of $\epsilon^{-1}(T,L)$.
This value is in good agreement with earlier estimates, based
on Coulomb gas simulations by
Saito and M\"uller-Krumbhaar,\cite{saito}
$T_{KT}=0.215$, and by Grest,\cite{Grest} $T_{KT}=0.220$.
In Fig.\ 2, we plot the resulting charge current density $J(E,L)$ versus
$E$, for several values of $L$, at the fixed temperature $T_{KT}=0.218$.
We see that the smaller $E$, the larger is the finite size effect
as $L$ varies.  From Eq.(\ref{eq:tcscal}) we see that
smaller values of $E$ probe larger length scales; correspondingly,
our statistical error increases as $E$ decreases.

To find the dynamic exponent $z$, we now fit the data of Fig.\ 2 to
the expanded scaling function of Eq.(\ref{eq:expscal}).
Since this expansion converges fastest for small values of the argument,
we restrict the data used in our fitting to those points where $EL\le 1$.
This corresponds to lattice sizes $L=6-14$, with $E=0.02-0.08$.
In Table I, we show the results of this fit, for several orders
of expansion, for various ranges of $L$.  Using the fourth order expansion
for lattice sizes $L=8-14$, we find $z=2.073\pm0.098$.
Using this $z$ in Eq.(\ref{eq:ivtc}), we get a non-linear response
$J\sim E^{3.073}$, consistent with
the prediction from the KT pair unbinding model, assuming the universal
jump in $\epsilon^{-1}(T_{KT})$ (see Eqs.(\ref{eq:je},\ref{eq:jump})).

In Fig.\ 3 we plot our data as $JL^z/E$ versus $EL$.  We see that the data
collapses onto a universal
curve representing the scaling function $\Phi(x,0,1)$.
The value $z=2.073$, obtained from the fitting,
is used in making the vertical axis, and
the solid line is drawn using the fitted values of the $\Phi_i$.
Although the agreement is reasonable, Table I does suggest some
potential problems.  The parameters $z\simeq 2$  and
$\Phi_0$, while remaining stable within the
estimated errors, both show a systematic increase as the smallest
$L_i$ used in the fit is increased.  $\Phi_2$, although
consistent within the different fits, shows a very
large estimated statistical error (other $\Phi_i$, although also
strongly fluctuating, seem too small to be significant in the fit).
Ideally, one would like to carry out these fits using increasingly
smaller values of $E$ than we have used here.  However, when $E$ becomes
small, equilibration times become large, and we were unable to get
accurate enough data to improve our fit.

Our results in Table I and Fig.\ 3 represent checks of scaling in the
small $EL$ limit.
We have also tried to check scaling in the large $EL$ limit.
Provided that $EL$ is sufficiently large that finite size effects are
small, we should be able to use Eq.(\ref{eq:dscal}) to collapse our
data onto two universal curves, given by $\Phi_+$ and $\Phi_-$ above
and below $T_{KT}$, by plotting $J\xi^z/E$ versus $E\xi/T$.  To do so,
we need an expression for the correlation length $\xi(T)$.
For the Kosterlitz-Thouless transition, asymptotically close to $T_{KT}$,
this form is,\cite{kt,Min}
\begin{equation}
\xi_\pm (T)\sim e^{C_\pm/|T-T_{KT}|^\nu}
\label{eq:xikt}
\end{equation}
where the subscripts $+$ and $-$ refer to behavior above and below $T_{KT}$
respectively, and for the KT transition, $\nu=1/2$.  Minnhagen and
Olsson\cite{olsson}
have argued that this true asymptotic form holds only in a narrow
critical region of about $5\%$ of $T_{KT}$.  Nevertheless, they also
indicate that Eq.(\ref{eq:xikt}) is a useful phenomenological
form for fitting over a wider temperature range
for $T>T_{KT}$, provided $C_+$ is
taken as a phenomenological parameter not necessarily equal to the true
asymptotic value.  We adopt this approach and use Eq.(\ref{eq:xikt})
with $C_\pm$ and $\nu$ as phenomenological parameters.

To carry out this large $EL$ check of scaling, we observe from
our data at $T_{KT}$ in Fig.\ 2, that finite size effects are
negligible provided we restrict the data to $L\ge 24$, $E\ge 0.06$.
Since this is true at $T_{KT}$, it should also certainly be true
for other values of $T$.  We therefore carry out simulations on
an $L=24$ lattice, for values $E\ge 0.14$.
Our results for $J(E,T)$ versus $E$, for various $T$ above and below
$T_{KT}$, are shown in Fig.\ 4. on a log-log scale.  Solid lines
with slopes of $1$ (for Ohmic behavior above $T_{KT}$), and $3$
(for critical behavior at $T_{KT}$) are shown for reference.
In Fig.\ 5, we try to collapse this data onto two universal curves
as discussed above, by finding the best choices for the parameters
$T_{KT}$, $z$, $C_\pm$, and $\nu$.  The results shown are for the
values $T_{KT}=0.218$, $z=2$ (consistent with our small $EL$
analysis), $\nu=1/2$ (consistent with the KT form), and $C_+=C_-=0.35$.
The collapse is reasonable, except for the smallest several values
of $T$ below $T_{KT}$.  This is most likely due to a failure of
the assumed form for $\xi(T)$,
Eq.(\ref{eq:xikt}), to be valid over such a large temperature range.

\subsection{$f=1/2$}
\label{sechalf}

In this section we carry out a similar analysis as in the previous
section, except applied to the $f=1/2$ Coulomb gas, which corresponds
to the ``fully frustrated" XY model.  As discussed at the end of
Section \ref{ktsec}, there are in principle two transitions in this
model: a insulator to metal transition at $T_{KT}$, and a charge
lattice melting transition at $T_M$.  It is $T_{KT}$
which corresponds to the transition from non-linear to linear $J-E$
characteristics (see Eq.(\ref{eq:je}) and following discussion).
The most recent equilibrium simulations of the
$f=1/2$ Coulomb gas model by one of us\cite{Leef12}
find that $T_{KT}\simeq 0.126$ is very close to,
but slightly below $T_M\simeq 0.1315$; the discontinuous jump
in $\epsilon^{-1}$ is
$1/\epsilon(T_{KT})T_{KT} \simeq 5.35$, larger than the universal KT
value of $4$ (see Eq.(\ref{eq:jump})).  This compares with earlier estimates
by Grest\cite{Grest} of $T_{KT}=0.129\pm0.002$,
with a jump $1/\epsilon(T_{KT})T_{KT} \simeq 4.88\pm0.31$.
Similar simulations on the fully frustrated XY model by
Ramirez-Santiago and Jos\'e\cite{Jose}
find $T_{KT}\approx T_M$ and a jump $1/\epsilon(T_{KT})T_{KT} \simeq 5.21$.

Fixing the temperature at $T_{KT}=0.126$, we show in Fig.\ 6 our results for
$J(E,L)$ versus $E$ for various $L$.  To extract the critical exponent
$z$ we fit this data to the expanded scaling function of Eq.(\ref{eq:expscal}).
We restrict\cite{note} the data used in our fitting
to those points where $EL\le 0.5$,
which corresponds to lattice sizes $L=6-14$, with $E=0.01-0.04$.
In Table II, we show the results of this fit.
Using the fourth order expansion
for lattice sizes $L=8-14$, we find $z=2.060\pm 0.124$.
As was found for $f=0$, the result $z\simeq 2$ is consistent with
a power law response of $J\sim E^3$.

In Fig.\ 7 we plot the data of Fig.\ 6
as $JL^z/E$ versus $EL$, and find fair agreement with the expected
collapse onto a universal curve. Our fitted value of $z=2.060$
is used in making the vertical axis, and
the solid line is drawn using the fitted values of the $\Phi_i$.
Although the agreement is reasonable, Table II again suggests some
potential problems.  As we found in Table I for the $f=0$ case,
now for $f=1/2$, the parameters $z$ and $\Phi_0$ show a systematic
increase as the smallest $L_i$ used in the fit is increased.  Now
however, this increase is more pronounced, and the fitted values remain
consistent with varying $L_i$ only within the outer limits of the
estimated statistical errors.  Furthermore, the higher $\Phi_i$ all
seem to be significant, and all have very large statistical error.
These observations make it unclear whether or not our data truly
represents the asymptotic scaling region of large $L$, small $EL$.

We have not attempted to check scaling for this $f=1/2$ model in the
large $EL$ limit.  The strong finite size effects seen in Fig.\ 6,
even comparing $L=24$ and $32$, means that we would have to go either
to larger lattice sizes (which are beyond our present computational
ability), or to temperatures sufficiently far from $T_{KT}$,
in order to reach the large $EL$ limit, for the values of $E$ we have studied.
Uncertainties in the correct form one should
take for $\xi(T)$, due in particular to the close proximity of the vortex
lattice melting transition at $T_M$ to the insulator to metal transition
at $T_{KT}$, would undoubtedly make such an analysis more complicated than
was the case for $f=0$.

\section{Conclusions}

To conclude, we have carried out steady-state driven diffusion Monte Carlo
simulations of the $2d$ lattice Coulomb gas, in order to compute the dynamic
exponent $z$, and hence obtain the non-linear response $J\sim E^a$, $a=z+1$,
at criticality.  This corresponds to the non-linear current-voltage
characteristic $V\sim I^a$ in a superconducting network at the
superconducting to normal transition.  We have analyzed our data
according to a finite-size scaling method
based on a new dynamic scaling conjecture by
Fisher {\it et al}.\cite{Fis}

For the $f=0$ model,
corresponding to a superconducting network in zero applied magnetic
field, our results agree with the familiar Kosterlitz-Thouless
pair unbinding model.  Our finite-size scaling analysis,
varying $L$ and $E$ at fixed $T=T_{KT}$, gives a value
of $z\simeq 2$, consistent with a power law response at $T_{KT}$ of $a=3$.
Our check of scaling in the infinite $L$ limit, where we vary
$T$ and $E$ for fixed $L$ large, shows fair agreement with the
Kosterlitz-Thouless model, but success is limited by our limited
knowledge of the form of the correlation length $\xi(T)$ outside the
narrow critical region.

For the $f=1/2$ model, corresponding to a superconducting network in
an applied magnetic field of one half flux quantum per unit cell,
we again find $z\simeq 2$.  This is consistent with the Kosterlitz-Thouless
pair unbinding result of Eq.(\ref{eq:je}) only if the discontinuous
jump in $\epsilon^{-1}(T_{KT})$ obeys the universal KT prediction of
Eq.(\ref{eq:jump}), ie. $1/\epsilon(T_{KT})T_{KT}=4$.  Equilibrium
simulations however indicate that for $f=1/2$, this jump
is non-universal, with $1/\epsilon(T_{KT})T_{KT}\simeq 5.35$.  Using this
value of $\epsilon$ in Eq.(\ref{eq:je}) yields the non-linear
response at $T_{KT}$ due to pair unbinding as, $J\sim E^a$, with $a=3.68$.
If pair unbinding were the dominant contribution to $J$, this would
imply a dynamic exponent of $z=a-1=2.68$.

It remains unclear
what is the source of this inconsistency.  It could be that the
analysis\cite{Leef12}
of $\epsilon^{-1}$ in the equilibrium simulations is in error;
this equilibrium analysis involves identifying leading logarithmic
corrections at $T_{KT}$ which may be hard to determine accurately.
Or it could be that our finite-size analysis of $z$ in this present
work is flawed; this possibility is indicated by the less than
satisfactory behavior of the fitted parameters (see Table II,
and discussion at the end of Section \ref{sechalf}) as
we vary the order of the fitting expansion, or the system sizes used
in the fit.  Or it could be that the result $z=2$ is a more general
property of such superconducting systems, which is independent of the
KT pair unbinding model; in this case one would expect that some
excitation other than pairs gives the dominant contribution
to $J$ at $T_{KT}$.  The natural guess for these other excitations
is the domain excitations of the ground state charge lattice.
However this would seem unlikely, if the charge lattice melting
transition $T_M$ is distinctly higher than $T_{KT}$, as equilibrium
simulations suggest.  Therefore, behavior in the $f=1/2$ model remains
an enigma, both from the equilibrium, and now from the
steady-state dynamic point of view.

\section*{Acknowledgments}

This work has been supported by U. S.  Department of Energy
grant DE-FG02-89ER14017, and in part by the Korea Science and Engineering
Foundation through the SRC program of SNU-CTP.

\begin{figure}
\caption{Ground state charge lattice for the $f=1/2$ Coulomb gas.
 A $+$ denotes the presence of a charge $n_i=+1$.}
\label{fig1}
\end{figure}

\begin{figure}
\caption{Charge current density $J(E,L)$ versus applied electric field $E$
for various square lattice sizes $L$, at fixed temperature $T_{KT}=0.218$,
for the $f=0$ model.
$10^6$ total MC passes were used to compute averages.}
\label{fig2}
\end{figure}

\begin{figure}
\caption{The finite size scaling behavior of the charge current density.
$JL^z/E$ versus $EL$ is plotted for various system sizes $L$ at fixed
temperature $T_{KT}=0.218$, for the $f=0$ model.
Symbols with error bars represent the MC data.
The solid line results from the fitting to Eq.(22)
using a fourth order expansion in $EL$,
and data from $L=8-14$ and $E=0.02-0.08$.
Data from $E=0.1$ and $0.12$ for each lattice size are included in the plot.
The fitted value of $z=2.073$ was used in making the vertical axis.
$10^6$ total MC passes were used to compute averages.}
\label{fig3}
\end{figure}

\begin{figure}
\caption{Charge current density $J(E,T)$ versus $E$ for various temperatures
$T$, for fixed system size $L=24$, for the $f=0$ model.
The solid lines from left to right have slopes $1$ and $3$ respectively,
illustrating Ohmic behavior above $T_{KT}$, and KT critical behavior at
$T_{KT}$. $10^6$ total MC passes were used to compute averages.}
\label{fig4}
\end{figure}

\begin{figure}
\caption{Test of the infinite size scaling relation Eq.(17)
for the $f=0$ model.
The data of Fig. 4 is replotted as $J\xi^z/E$ versus $E\xi/T$, using
the form of Eq.(23) to determine $\xi (T)$.  Parameters $z$,
$T_{KT}$, $\nu$, and $C_\pm$ are varied, to get the best collapse of
the data onto two distinct scaling functions $\Phi_\pm$, above and
below $T_{KT}$.}
\label{fig5}
\end{figure}

\begin{figure}
\caption{Charge current density $J(E,L)$ versus applied electric field $E$
for various square lattice sizes $L$, at fixed temperature $T_{KT}=0.126$,
for the $f=1/2$ model.
$10^6$ total MC passes were used to compute averages.}
\label{fig6}
\end{figure}

\begin{figure}
\caption{The finite size scaling behavior of the charge current density.
$JL^z/E$ versus $EL$ is plotted for various system sizes $L$ at fixed
temperature $T_{KT}=0.126$, for the $f=1/2$ model.
Symbols with error bars represent the MC data.
The solid line results from the fitting to Eq.(22)
using a fourth order expansion in $EL$,
and data from $L=8-14$ and $E=0.01-0.04$.
Data from $E=0.05$ and 0.06 for each lattice size are included in the plot.
The fitted value of $z=2.060$ was used in making the vertical axis.
$10^6$ total MC passes were used to compute averages.}
\label{fig7}
\end{figure}

%
\begin{table}
\caption{Results of the fitting of $J$ to an expansion of the scaling
function in powers of $EL$, as in Eq.(22).
The data of Fig. 2 for the $f=0$ model is used.  The first column shows
the range of system sizes $L_i-L_f$ which are included in the
particular fit.  The following columns give the
fitted parameters.  The last column
is the $\chi^2$ error of the fit.  For each sequence of $L_i-L_f$,
the first row gives the value of the fitted parameter, while the
second row gives the estimated error.}
\label{tab1}
\begin{tabular}{|l|r|l|l|l|r|r|r|}
\multicolumn{1}{|c|}{$L_i-L_f$} &
 \multicolumn{1}{c|}{$z$} &
 \multicolumn{1}{c|}{$\Phi_0$} &
 \multicolumn{1}{c|}{$\Phi_1$} &
 \multicolumn{1}{c|}{$\Phi_2$} &
 \multicolumn{1}{c|}{$\Phi_3$} &
 \multicolumn{1}{c|}{$\Phi_4$} &
 \multicolumn{1}{c|}{$\chi^2_{fit}$} \\ \hline
\multicolumn{8}{|c|}{1st order expansion}\\ \hline
\multicolumn{1}{|c|}{$6-14$}
  &1.9831    &0.1017    &0.0935   &          &         &   &8.8463 \\
  &0.0590    &0.0106    &0.0215   &          &         &   &  \\
\multicolumn{1}{|c|}{$8-14$}
  &2.0297    &0.1070    &0.1148   &          &         &   &6.5589 \\
  &0.1097    &0.0261    &0.0459   &          &         &   &  \\
\multicolumn{1}{|c|}{$10-14$}
  &2.0777    &0.1119    &0.1415   &          &         &   &5.0160 \\
  &0.1382    &0.0393    &0.0535   &          &         &   & \\ \hline
\multicolumn{8}{|c|}{2nd order expansion}\\ \hline
\multicolumn{1}{|c|}{$6-14$}
  &2.0067    &0.1281    &0.0137   &0.0734    &         &   &4.8485 \\
  &0.0663    &0.0253    &0.0740   &0.0733    &         &   & \\
\multicolumn{1}{|c|}{$8-14$}
  &2.0747    &0.1541    &0.0040   &0.0995    &         &   &3.6025 \\
  &0.1163    &0.0524    &0.1058   &0.1038    &         &   & \\
\multicolumn{1}{|c|}{$10-14$}
  &2.1203    &0.1657    &0.0195   &0.1053    &         &   &2.9278 \\
  &0.1238    &0.0677    &0.1284   &0.1079    &         &   & \\ \hline
\multicolumn{8}{|c|}{3rd order expansion}\\ \hline
\multicolumn{1}{|c|}{$6-14$}
  &2.0060    &0.1268    &0.0192   &0.0631    &0.0052   &   &4.8474 \\
  &0.0953    &0.0345    &0.1114   &0.1335    &0.1307   &   & \\
\multicolumn{1}{|c|}{$8-14$}
  &2.0725    &0.1520    &0.0132   &0.0835    &0.0077   &   &3.6012 \\
  &0.1075    &0.0373    &0.0940   &0.1105    &0.1159   &   & \\
\multicolumn{1}{|c|}{$10-14$}
  &2.1208    &0.1669    &0.0131   &0.1174   &-0.0065   &   &2.9228 \\
  &0.1220    &0.0525    &0.1906   &0.2496    &0.1196   &   & \\ \hline
\multicolumn{8}{|c|}{4th order expansion}\\ \hline
\multicolumn{1}{|c|}{$6-14$}
  &2.0101    &0.1219    &0.0076   &0.0790   &0.0068   &-0.0061 &4.8552 \\
  &0.0722   &0.0290    &0.0693   &0.1261   &0.0545   & 0.0170 & \\
\multicolumn{1}{|c|}{$8-14$}
  &2.0727    &0.1517    &0.0144   &0.0789   &0.0151   &-0.0040 &3.5949 \\
  &0.0984    &0.0284    &0.0928   &0.1073   &0.0694   & 0.0491 & \\
\multicolumn{1}{|c|}{$10-14$}
  &2.1175    &0.1638    &0.0227   &0.0966   &0.0115   &-0.0064 &2.9201 \\
  &0.0980    &0.0304    &0.0825   &0.1252   &0.0455   & 0.0400 & \\
\end{tabular}
\end{table}

\begin{table}
\caption{Results of the fitting of $J$ to an expansion of the scaling
function in powers of $EL$, as in Eq.(22).
The data of Fig. 6 for the $f=1/2$ model is used.  The first column shows
the range of system sizes $L_i-L_f$ which are included in the
particular fit.  The following columns give the
fitted parameters.  The last column
is the $\chi^2$ error of the fit.  For each sequence of $L_i-L_f$,
the first row gives the value of the fitted parameter, while the
second row gives the estimated error.}
\label{tab2}
\begin{tabular}{|l|r|r|r|r|r|r|r|}
\multicolumn{1}{|c|}{$L_i-L_f$} &
 \multicolumn{1}{c|}{$z$} &
 \multicolumn{1}{c|}{$\Phi_0$} &
 \multicolumn{1}{c|}{$\Phi_1$} &
 \multicolumn{1}{c|}{$\Phi_2$} &
 \multicolumn{1}{c|}{$\Phi_3$} &
 \multicolumn{1}{c|}{$\Phi_4$} &
 \multicolumn{1}{c|}{$\chi^2_{fit}$} \\ \hline
\multicolumn{8}{|c|}{1st order expansion}\\ \hline
\multicolumn{1}{|c|}{$6-14$}
  &1.8976  &0.7788 & 1.3337 &        &        &          &14.2661 \\
  &0.0869  &0.1134 & 0.6810 &        &        &          & \\
\multicolumn{1}{|c|}{$8-14$}
  &1.9556  &0.8891 & 1.5631 &        &        &          &12.3852 \\
  &0.0954  &0.1659 & 0.8506 &        &        &          & \\
\multicolumn{1}{|c|}{$10-14$}
  &1.9894  &0.9327 & 1.7894 &        &        &          &10.5980 \\
  &0.0591  &0.0573 & 0.6560 &        &        &          & \\ \hline
\multicolumn{8}{|c|}{2nd order expansion}\\ \hline
\multicolumn{1}{|c|}{$6-14$}
  &1.9158  &0.9092 & 0.6484 & 1.1746 &        &          &11.2051 \\
  &0.0921  &0.1706 & 0.6392 & 0.6416 &        &          & \\
\multicolumn{1}{|c|}{$8-14$}
  &2.0622  &1.4525 &-0.2034 & 3.4457 &        &          & 4.5429 \\
  &0.1233  &0.5364 & 0.5575 & 2.7420 &        &          & \\
\multicolumn{1}{|c|}{$10-14$}
  &2.1514  &1.8426 &-0.5706 & 4.8902 &        &          & 3.0757 \\
  &0.0680  &0.2814 & 0.9038 & 1.7059 &        &          & \\ \hline
\multicolumn{8}{|c|}{3rd order expansion}\\ \hline
\multicolumn{1}{|c|}{$6-14$}
  &1.9128  &0.8566 & 1.2388 &-0.9075 & 2.0827 &          &11.0736 \\
  &0.0938  &0.2125 & 1.0552 & 3.6274 & 3.5002 &          & \\
\multicolumn{1}{|c|}{$8-14$}
  &2.0602  &1.3639 & 0.7936 & 0.0519 & 3.3454 &          & 4.3721 \\
  &0.1246  &0.3867 & 1.8768 & 4.4841 & 6.9044 &          & \\
\multicolumn{1}{|c|}{$10-14$}
  &2.1581  &1.6954 & 1.5878 &-2.4100 & 7.3724 &          & 2.5403 \\
  &0.0699  &0.2873 & 1.0701 & 1.8182 & 5.3878 &          & \\ \hline
\multicolumn{8}{|c|}{4th order expansion}\\ \hline
\multicolumn{1}{|c|}{$6-14$}
  &1.9125  &0.8636 & 1.0860 & 0.0985 &-0.4950 &2.1906    &11.0117 \\
  &0.0896  &0.1444 & 0.9954 & 1.7584 & 2.9752 &3.0877     & \\
\multicolumn{1}{|c|}{$8-14$}
  &2.0597  &1.3482 & 0.8797 & 0.3090 & 1.4246 &2.2639     &4.3701 \\
  &0.1238  &0.4363 & 0.9554 & 1.1924 & 2.0356 &5.9653     & \\
\multicolumn{1}{|c|}{$10-14$}
  &2.1554  &1.6883 & 1.3351 &-0.2117 & 1.0093 &5.6638     &2.5421 \\
  &0.0821  &0.3368 & 1.3246 & 1.9165 & 1.6536 &6.1369     & \\
\end{tabular}
\end{table}


\begin{references}
\bibitem{note}
We chose a smaller limit for $EL$, as compared to our fits for
$f=0$, because (as we will see in Table II) even with this smaller limit, the
higher order coefficients $\Phi_i$ are more significant for $f=1/2$
than for $f=0$.
\end{references}
\end{document}